\documentclass[11pt]{article}
\usepackage{authblk} 
\usepackage{graphicx}
\graphicspath{ {/home/user/images/} }
\usepackage{dcolumn}
\usepackage{bm}
\usepackage{amsmath}
\usepackage{abstract}
\usepackage{xurl}
\usepackage[super,comma,sort&compress]{natbib}
\usepackage{abstract}

\usepackage{lipsum}

\title{\vspace{-3cm}\textbf{{Absolute Flux Calibrations for the Nancy Grace Roman Space Telescope Coronagraph Instrument}}}
\author[1,2]{Lindsey Payne}
\author[2]{Robert T. Zellem}
\author[2]{Marie Ygouf}
\author[1]{Bruce Macintosh}
\affil[1]{Department of Physics, Stanford University}
\affil[2]{Jet Propulsion Laboratory – California Institute of Technology}
\date{\today}

\begin{document}

\maketitle
    \begin{abstract}
    The Nancy Grace Roman Space Telescope's (Roman) Coronagraph Instrument is a technology demonstration equipped to achieve flux contrast levels of up to 10$^{-9}$. This precision depends upon the quality of observations and their resultant on-sky corrections via an absolute flux calibration (AFC). Our plan utilizes 10 dim and 4 bright standard photometric calibrator stars from Hubble Space Telescope's (HST) CALSPEC catalog to yield a final AFC error of 1.94\% and total observation time of $\sim$22 minutes. Percent error accounts for systematic uncertainties (filters, upstream optics, quantum efficiency) in Roman component instrumentation along with shot noise for a signal to noise ratio (SNR) of 500.
    \end{abstract}


\maketitle
\section{Introduction}
NASA’s next space-based observatory will be launched in no later than 2027 to carry out groundbreaking work in direct imaging and exoplanet detection. Roman contains the Wide Field Instrument which will conduct an expansive microlensing survey and probe the cosmos for information regarding dark energy and infrared astrophysics. The 2.4-meter observatory also hosts the revolutionary Coronagraph Instrument: an advanced Technology Demonstration in high-contrast exoplanet imaging \cite{Roman}. 
 
The Coronagraph affords photometry, polarimetry, and spectroscopy across 4 bandpasses. Photometry, or imaging, and polarimetry are centered at 575 nm and 835 nm in Band 1 (10\% bandwidth) and Band 4 (15\% bandwidth), respectively \cite{params}. Slit spectroscopy is centered at 660 nm in Band 2 and 730 nm in Band 3 (6\% bandwidth for both), with a resolution of R$\sim$50. Deformable mirrors (DMs), an adaptive optics mechanism in the Coronagraph, correct for wavefront aberrations in real-time. This is the first time DMs will be used in a space-based observatory \cite{DMs}. Thanks to that technology, the Coronagraph will be capable of imaging reflective planets at visible wavelengths for the first time as well.
		
Coronagraphs are used to block the light of a host star such that its orbiting planet can be directly imaged. The Roman Coronagraph intends to achieve record contrast levels between dim planets and their bright host stars, $F_{ratio} = \frac{F_{planet}}{F_{star}}$, at the level of at least 10$^{-7}$ and up to 10$^{-9}$. Achieving this contrast goal would afford new and important observations of very dim and/or distant exoplanets that can be further studied by astronomers or astrophysicists interested in habitability, planetary formation, etc. 

These challenging contrast levels require high precision in the measured flux of a target system – a dim planet and its bright host star. Flux measurement precision is limited by an instrument's sources of systematic uncertainty, as well as random photon noise. All photometric data must be absolute-flux-calibrated, using standard stars, to mitigate these errors as well as prepare for unanticipated uncertainties post-launch. Such uncertainties might include variation in our filter transmissivities due to condensation from the space environment, for example.

The AFC not only removes unwanted throughput contributions and interferences from our flux measurements, but is used to convert these measurements from detector units of photoelectrons to physical units of apparent magnitude. This allows our data to be in a useful metric space for additional scientific interpretation. However, since the AFC only provides a localized measurement of Roman’s total effective throughput, it must be combined with a flatfield image to account for full focal plane gain offsets. The Coronagraph will not necessarily know the position angle and separation of a planet relative to its host star before hand, thereby requiring this calibration of the larger focal plane, rather than a localized region of interest. These calibrations will also enable the correction for the total effective throughput of the entire optical beam train (the Roman observatory and the Coronagraph itself) and, if needed, its spectral response.

\maketitle
\section{Methods}
Here, we detail our method of simulating an observation with our Coronagraph instrument. We want to predict what we will see and set up our post-launch AFC accordingly so the mission can stay within its allocated AFC error budget of 2.9\%. This is carried out using a robust Python script that loads filter data, generates spectral models, simulates an observation for a chosen host-planet system, and outputs a percent error between the expected flux ratio and simulated one (Eqn.4), as well as a percent error on the flux ratio after it is calibrated with two standard stars, 1 dim and 1 bright (Eqn.5). Our script leverages a synthetic photometry package, called synphot, to help simulate our chosen spectra using Roman filter data \cite{synphot}.

\begin{flalign}
C_{i} = \frac{F_{s,i}}{F_{s,truth}}
\end{flalign}

\begin{flalign}
F_{C, i} = F_{t, i} * C_{i}
\end{flalign}

\begin{flalign}
F_{C,ratio} = \frac{F_{C,planet}}{F_{C,star}}
\end{flalign}

\begin{flalign}
\sigma_{F_{raw,ratio}} = \mid \frac{F_{i,ratio} - F_{truth,ratio}}{F_{truth,ratio}} \mid * 100
\end{flalign}

\begin{flalign}
\sigma_{F_{C,ratio}} = \mid \frac{F_{C,ratio} - F_{truth,ratio}}{F_{truth,ratio}} \mid * 100
\end{flalign}
 
Standard spectra are loaded from fits files in HST'S CALSPEC database \cite{CALSPEC}. We chose CALSPEC stars because of Hubble’s flight heritage, model accessibility, spectral type variety, and detailed grid of systematic errors. Furthermore, HST observes in similar wavelengths to the Coronagraph. Target spectra are generated using the PHOENIX stellar modeling software \cite{PHOENIX}. Target exoplanet spectra are either loaded from self-luminous models or generated using the target spectrum and an albedo function for Jupiter \cite{LACY}. 
		
We select which standard stars to pass through our simulations according to their observational visibility (Fig.1 and Fig. 2), specific to the Coronagraph, as well as their brightness up to a saturation level for a specific filter \cite{Exposure}. The ND filter is required to observe the relatively bright unocculted flux of the host star, or bright standard star, without saturating, and has two optical descriptions (OD): 4.75 and 2.25, a $10^{-4.75}$ and $10^{-2.25}$ flux reduction, respectively. Dim standards are observed with the color filter only, in contrast to bright standards that are observed with both the color and ND filters. These standard star selection criteria of visibility and brightness help us optimize observation time for the Roman Coronagraph's AFC.

\begin{figure}[!h]
\includegraphics[scale=0.57]{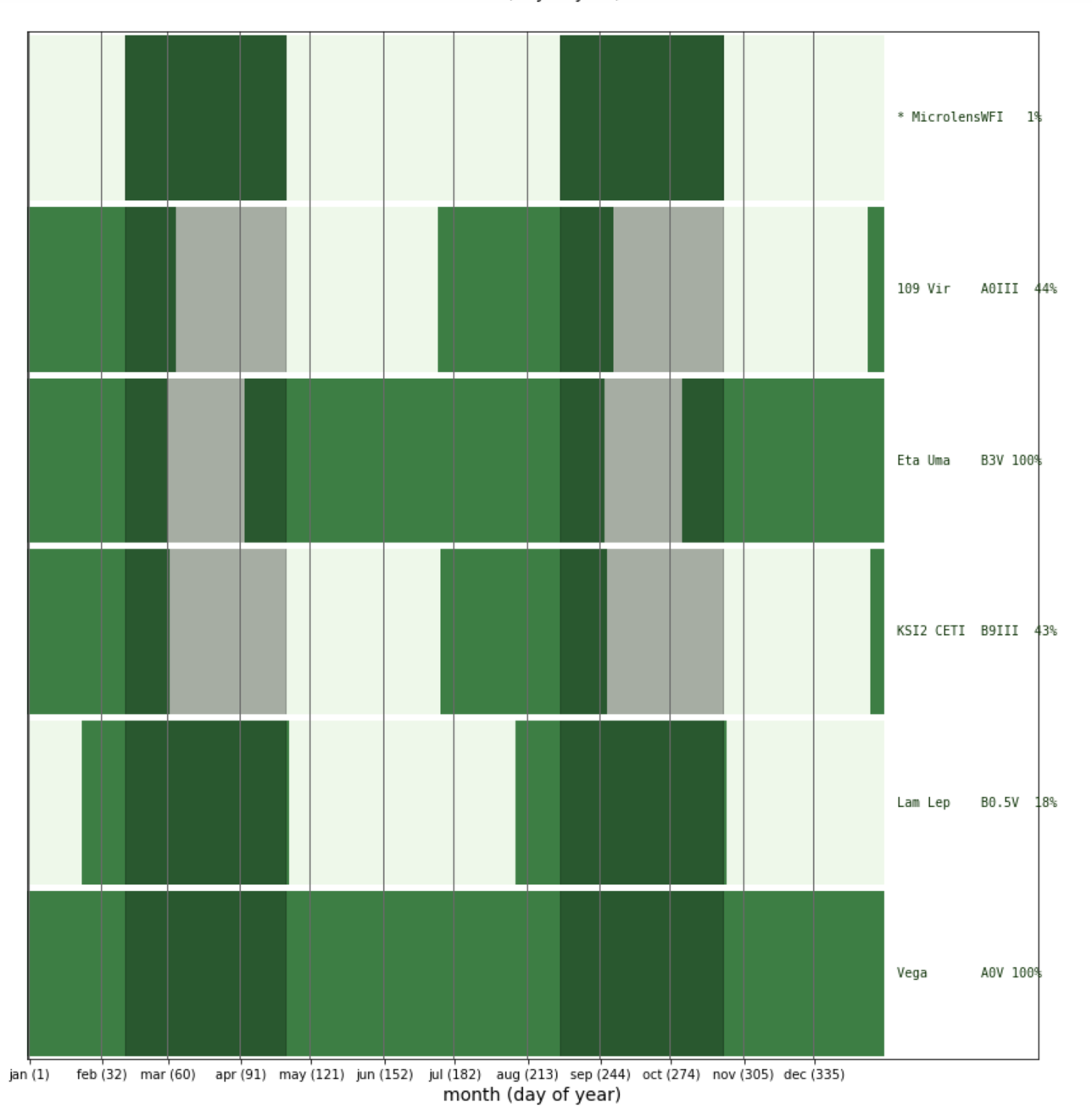}
\caption{\textbf{Bright CALSPEC standard star visibilities per Roman constraints.}}{\label{fig:epsart} Displays visibility percentages of bright standard stars according to identifier name, alphabetically, from top to bottom. Percentages indicate the Roman Coronagraph’s ability to observe the star outside of the mission’s designated periods for microlensing – February to April and August to October. Dark green represents microlensing time, medium green means that the star will be visible, and light green indicates the star will not be visible at that time.}
\end{figure}

\begin{figure}[!h]
\includegraphics[scale=0.57]{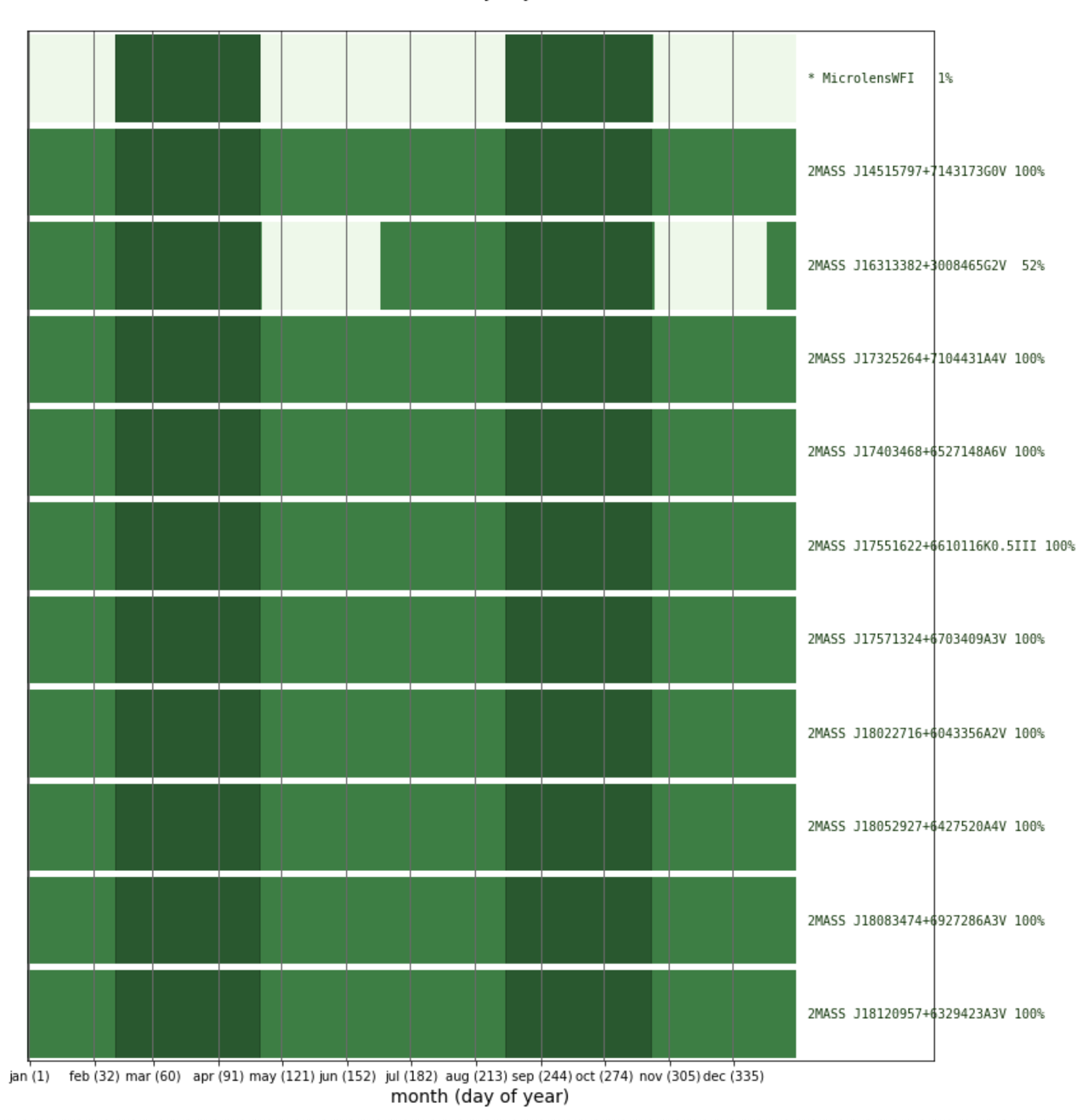}
\caption{\textbf{Dim CALSPEC standard star visibilities per Roman constraints.}}{\label{fig:epsart} Displays visibility percentages of dim standard stars according to identifier name, alphabetically, from top to bottom. Percentages indicate the Roman Coronagraph’s ability to observe the star outside of the mission’s designated periods for microlensing – February to April and August to October. Dark green represents microlensing time, medium green means that the star will be visible, and light green indicates the star will not be visible at that time.}
\end{figure}

Our script leverages a Monte Carlo (MC) sampling method. We sample 5 different sources of systematic uncertainty and 1 source of random uncertainty when simulating an observation with the Roman Coronagraph system. This ensures a conservative error propagation of the 6 sources: color filter transmission ($\pm$ 0.04\% error on vendor specifications), neutral density filter (ND) transmission for bright stars only ($\pm$ 3.83\%), other upstream Roman optics ($\pm$ 20\%), quantum efficiency (QE) of the electron multiplying charge-coupled device ($\pm$ 2.0\%), CALSPEC standard star uncertainties, and shot noise, respectively. CALSPEC spectral uncertainties are derived from a covariance matrix referenced in Bohlin et al. 2014 and applied to each standard spectrum at the appropriate wavelength bin \cite{BOHLIN2014}. Shot noise obeys the Poisson distribution; we set an SNR of 500 for all simulated observations to minimize the contribution of shot noise to our error budget. 

 
With each iteration, we build up distinct distributions of simulated flux values for the chosen target star, target planet, bright standard star, and dim standard star. Outside of the MC loop, we compute the true flux value of each object by integrating its known spectrum over a chosen bandpass. We then take the distribution of MC simulated flux values for our bright standard star, $F_{s,i}$, and divide each by the standard's known flux value, $F_{s,truth}$, to get a resultant distribution of scaling factors, or correction terms (Eqn.1). This is repeated for the dim standard star as well. We then adjust our simulated target planet and target star fluxes, $F_{t,i}$,by multiplying each distribution by its respective distribution of correction terms – dim and bright (Eqn.2). This leaves us with an absolute-flux-corrected distribution of planet and host star flux measurements, $F_{C,planet}$ and $F_{C,star}$, that may be turned into a flux ratio (Eqn.3) and used to analyze our calibration performance (Eqn.5). 

Here, standard stars and AFC errors are reported for Band 1 photometry. Distribution values are compared to their truth value, $F_{truth \; , \; ratio}$, with no embedded uncertainties. Comparisons inform the output errors referenced above and defined below: a raw flux ratio error (Eqn.4) and an AFC-corrected flux ratio error (Eqn.5) for a given target-standard pairing.

\maketitle
\section{Results \& Conclusions}

The Coronagraph team has identified 8 potential star-planet systems that could be observed during the Roman Coronagraph Technology Demonstration phase. Through thorough testing, parameter adjustment, and error analysis, we have deduced that the optimal set of standard stars to use in the AFC includes 10 dim (Eqn.6) and 4 bright (Eqn.7) CALSPEC standard stars. Testing determined that 10,000 MC iterations and an SNR of 500 yield statistically significant results with minimal variation between runs of the script.

\begin{figure}[!h]
\includegraphics[scale=0.48]{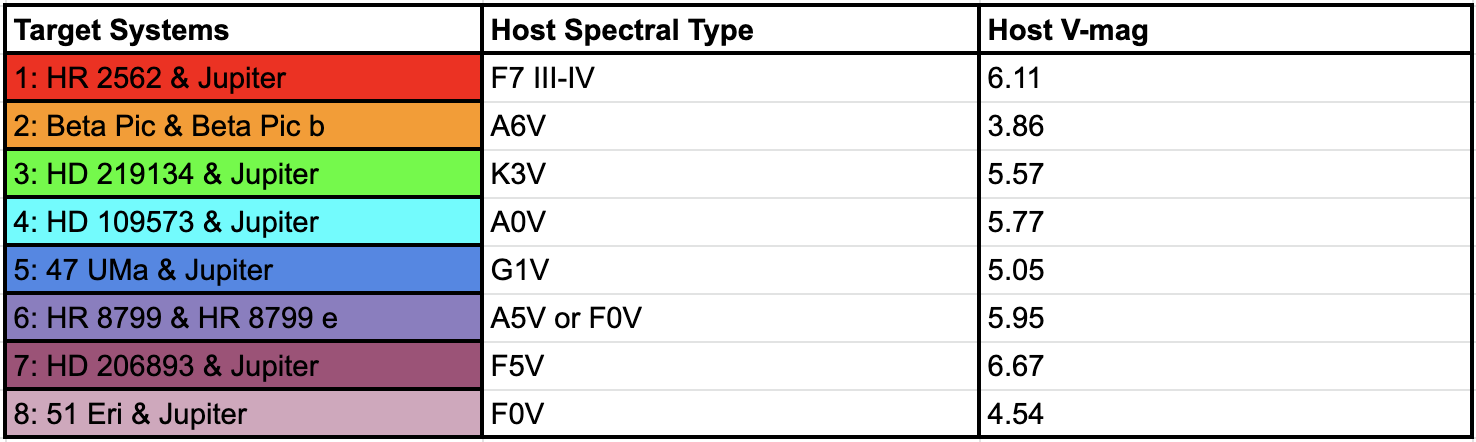}
\caption{\label{fig:epsart} Table of possible Coronagraph mission target systems. Left column states the system number, host star name, and exoplanet to be modeled. Right column states the host star's spectral type extracted from SIMBAD.}
\end{figure}

Taking into account exposure time alone, it will take 21 minutes and 39 seconds to observe these 14 total stars. Note that this estimate does not include slewing nor other observational overheads, but all chosen standards sit in the northern celestial hemisphere which will help to minimize this time. Combining the maximum output error between runs (Eqn.8) for each standard star yields a current best estimate (CBE) of 1.944\% for the Coronagraph’s total AFC uncertainty (from Eqn.5). This maintains a 33\% margin on the allocated error budget.

\begin{@twocolumnfalse}
    \begin{figure}[!h]
    \includegraphics[scale=0.3]{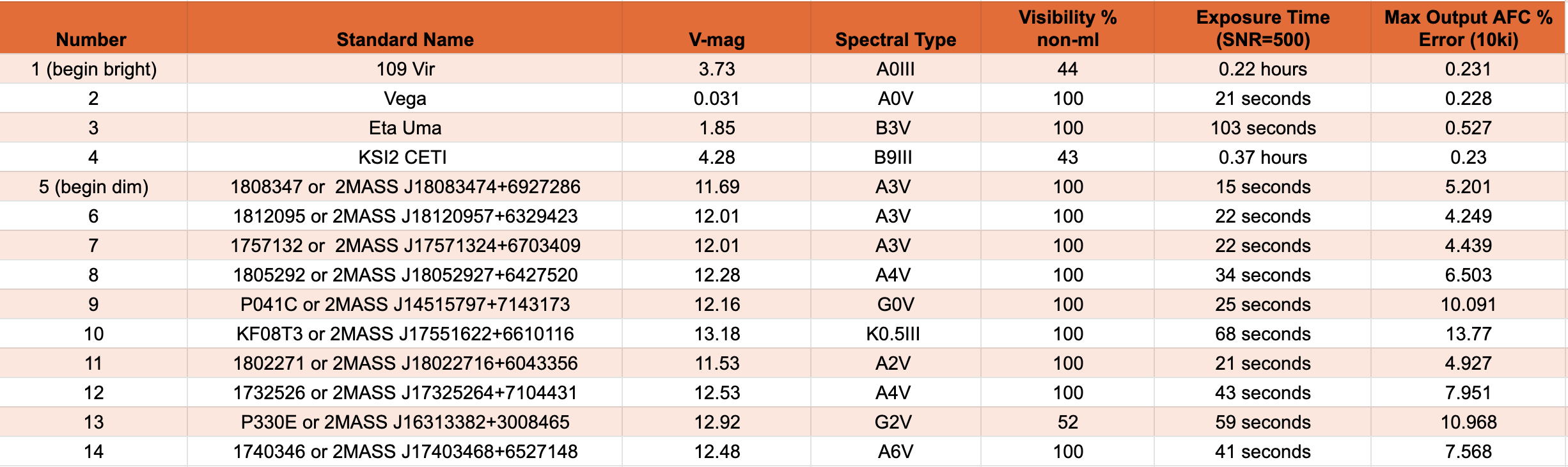}
    \caption{\label{fig:epsart} Table of chosen standard star names, apparent V-band magnitudes, spectral types, visibility percentages in the non-microlensing (non-ml) period, required exposure times, and maximum (across all 8 potential target systems) percent errors in the calibrated flux ratio from 10,000 iteration MC simulations.}
    \end{figure}
\end{@twocolumnfalse}

This error combination, and reduction, is motivated by the James Webb Space Telescope (JWST) AFC plan which bins observations from multiple CALSPEC standard stars to reduce the total uncertainty on the AFC \cite{JWST}.

\begin{flalign}
     \sigma_{dim} = \frac{1}{\sqrt{\left(\frac{1}{\sigma_{1}}\right)^{2}+\left(\frac{1}{\sigma_{2}}\right)^{2}+ ... +\left(\frac{1}{\sigma_{10}}\right)^{2}}}
\end{flalign}

\begin{flalign}
     \sigma_{bright} = \frac{1}{\sqrt{\left(\frac{1}{\sigma_{1}}\right)^{2}+\left(\frac{1}{\sigma_{2}}\right)^{2}+ ... +\left(\frac{1}{\sigma_{4}}\right)^{2}}}
\end{flalign}

\begin{flalign}
    \sigma_{combined} = 10^{-9}\cdot\sqrt{\left(\frac{\sigma_{dim}}{10^{-9}}\right)^{2}+\left(\frac{\sigma_{bright}}{1}\right)^{2}}
\end{flalign}

Each additional star reduces the total error on the AFC. We use fewer stars than JWST in order to maximize the Coronagraph efficiency, while still achieving our uncertainty allocation with margin. Specifically, our observing campaign will use 10 dim and 4 bright CALSPEC standard stars to meet our 2.9\% allocation.

We find the dominant source of error in raw ratio measurements is the ND filter. In contrast, the dominant source of error in calibrated ratio measurements is CALSPEC systematic uncertainties on the standard spectra. This source disproportionately affects our dim standard candidates. Spectral type matching, between target object and standard star, impacts errors negligibly. There only exists a correlation between apparent magnitude and error: ultimately, brighter is better due to better calibrated CALSPEC uncertainties at these larger fluxes. 

 
Future work includes finding a unique and optimal set of standard stars for observing modes 2 and 3.

\maketitle
\section{Acknowledgements}
    Thank you to the entire Roman Coronagraph team, as well as Ralph Bohlin and Karl Gordon from the Space Telescope Science Institute for their insights.

This research was carried out at the Jet Propulsion Laboratory, California Institute of Technology, under a contract with the National Aeronautics and Space Administration (80NM0018D0004).

\end{document}